\documentclass[aps,prl,showpacs,superscriptaddress,amsmath,amssymb,floatfix,twocolumn]{revtex4}
\usepackage{graphicx}
\usepackage{hyperref}
\usepackage{url}

\begin{document}

\widetext

\title{Electron interferometer formed with a scanning probe tip and quantum point contact}

\author{M. P. Jura} \affiliation{Department of Applied Physics, Stanford University,
Stanford, California 94305, USA}
\author{M. A. Topinka} \affiliation{Department of Physics, Stanford University, Stanford, California 94305, USA} \affiliation{Department of Materials Science \& Engineering, Stanford University, Stanford, California 94305, USA}
\author{M. Grobis$^*$} \affiliation{Department of Physics, Stanford University, Stanford, California 94305, USA}
\author{L. N. Pfeiffer} \affiliation{Bell Labs, Alcatel-Lucent, Murray Hill, New Jersey 08544, USA}
\author{K. W. West} \affiliation{Bell Labs, Alcatel-Lucent, Murray Hill, New Jersey 08544, USA}
\author {D. Goldhaber-Gordon$^\dag$} \affiliation{Department of Physics, Stanford University, Stanford, California 94305, USA}


\begin{abstract}
We show an electron interferometer between a quantum point contact (QPC) and a scanning gate microscope (SGM) tip in a two-dimensional electron gas.  The QPC and SGM tip act as reflective barriers of a lossy cavity; the conductance through the system thus varies as a function of the distance between the QPC and SGM tip.  We characterize how temperature, electron wavelength, cavity length, and reflectivity of the QPC barrier affect the interferometer.  We report checkerboard interference patterns near the QPC and, when injecting electrons above or below the Fermi energy, effects of dephasing.
\end{abstract}


\maketitle

When electronic device dimensions become smaller than the electron coherence length, the wavelike nature of electrons becomes critical to understanding device operation and provides opportunities to build devices taking advantage of quantum properties.  Interferometers have been used to study electron interference in systems such as carbon nanotubes \cite{Park-FP} and GaAs two-dimensional electron gases (2DEGs) \cite{Interferometers}.  Direct spatial visualizations of interference effects appear as fringes in images of electron flow in 2DEGs taken by scanning gate microscopy (SGM) \cite{Topinka-Science,Topinka-Nature} and can give information about the local potential \cite{LeRoy-LocalDensity,Jura-NaturePhysics}.

Fringes appear in SGM images when different reflected electron paths interfere.  Previously observed fringes were due to an interferometer similar to a Michelson interferometer. One path of the interferometer is created by an SGM tip and the other path by impurities \cite{Topinka-Science,Topinka-Nature,LeRoy-LocalDensity} or a reflector gate \cite{LeRoy-Interferometer}.  However, fringes were not seen in high-mobility samples with a low density of impurities \cite{Jura-NaturePhysics}.  In this Rapid Communication we report interference fringes in SGM images taken in one of the same high-mobility samples at lower temperatures.  Now multiple interfering paths are created by the SGM tip, similar to the movable reflector of a Fabry-P\'{e}rot interferometer.  We report spatial interference patterns different from those previously observed.  The recovery of fringes in clean samples at lower temperatures allows us to spatially probe phase coherent properties, such as local phase and electron wavelength, and we demonstrate how these may be useful for studying dephasing.  Furthermore, the understanding of interference fringes which we achieve is necessary before using fringes of a different origin to measure electron interactions in nanostructures \cite{Freyn-Fringes}.

Thermal averaging limits how far from a coherent source interference effects can be observed, even if each individual electron is still coherent \cite{Heller-ThermalWP}.  Electrons around the Fermi energy with a spread in energies comparable to the thermal energy are involved in transport and become out of phase after traveling a thermal length, $L_T = h^2/2 \pi m \lambda_F k_B T$ (where $\lambda_F$ is the Fermi wavelength and $m$ is the effective electron mass).  In the simplest case of 2 paths interfering, in addition to the requirement that each path be shorter than the coherence length, the difference in length between the two paths must be shorter than $L_T$ for interference to be visible.

\begin{figure}
\begin{center}
\includegraphics[width=3.375in]{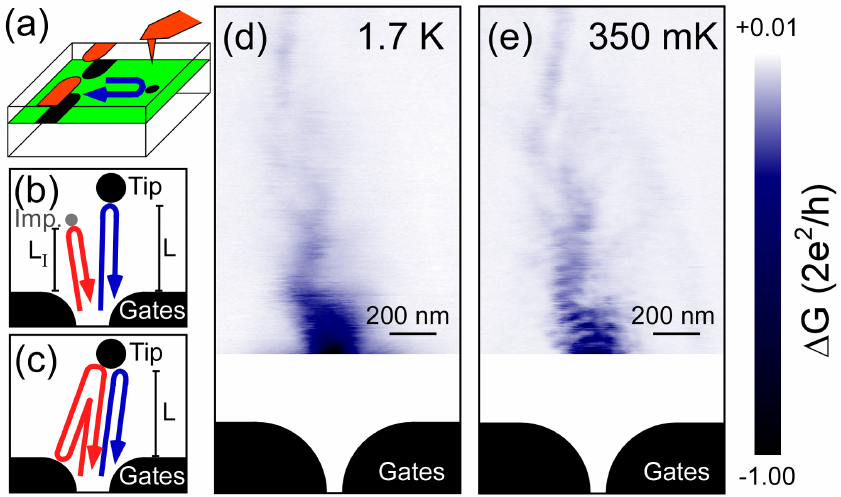}
\end{center}
\caption[Mechanisms for interference fringe formation and appearance of fringes at low temperature]{Mechanisms for interference fringe formation and appearance of fringes at low temperature.  (a) Schematic of imaging technique showing the 2DEG (green), as well as surface gates and metallic tip (orange) creating depletion regions (black) in the 2DEG below.  When above an area of high electron flow, the tip scatters electrons back through the QPC (blue path) and reduces the measured conductance $\Delta G$.  (b) Fringe mechanism 1: an impurity (gray) near the 2DEG acts as a hard scatterer and creates a second electron path to the QPC (red).  (c) Fringe mechanism 2: similar to an optical Fabry-P\'{e}rot interferometer.  (d) Electron flow at $1.7 \ \mathrm{K}$ shows no interference fringes.  Mechanism 1 does not occur, and the sample is too warm for mechanism 2 to be visible.  (e) Electron flow at $350 \ \mathrm{mK}$ showing interference fringes at the bottom of the image due to mechanism 2.}
\end{figure}

We image electron flow emanating from a quantum point contact (QPC) in a GaAs 2DEG using an SGM situated in a $^3$He cryostat with a base temperature of $350 \ \mathrm{mK}$ \cite{EPAPS}.  The 2DEG (same as sample C in Reference \cite{Jura-NaturePhysics}) has density $1.5 \ \times \ 10^{11} \ \mathrm{cm}^{-2}$ and mobility $4.4 \ \times \ 10^{6} \ \mathrm{cm}^2/\mathrm{V\,s}$ at $4.2 \ \mathrm{K}$.  We measure the conductance across a split-gate QPC \cite{QPCs} as we scan a metallic SGM tip ${\sim}30 \ \mathrm{nm}$ above the surface of the sample, as depicted in Fig. 1(a).  Application of negative voltage to the tip creates a depletion disk in the 2DEG below.  When the depletion disk is in a region of high electron flow, it scatters electrons back through the QPC, reducing the measured conductance [path depicted in blue in Fig. 1(a)-(c)].  By scanning the tip and recording the change in conductance, $\Delta G$, we can thus map electron flow \cite{Topinka-Science}.  Unless stated otherwise, the QPC conductance is set to $2e^2/h$, the middle of the first plateau, in the absence of a tip-induced depletion disk.

Interference fringes can appear in images of electron flow if other backscattering paths exist in addition to the path probed by the tip.  We consider two distinct mechanisms with very different temperature dependences.  Mechanism 1 is depicted in Fig. 1(b): hard scatterers (depicted in gray created, for example, by impurities near the 2DEG) reflect an appreciable amount of electron flux directly backward through the QPC (red path).  As the tip changes position, the length of the blue path changes but the length of the red path does not, causing a full cycle of interference as the tip moves away from the QPC by $\lambda_F/2$ \cite{Heller-Branching}.  Interference fringes are visible when the tip-to-QPC distance, $L$, is within $L_T/2$ of the impurity-to-QPC distance, $L_{I}$, (i.e. $L_{I} - L_T/2 < L < L_{I} + L_T/2$) \cite{Heller-ThermalWP}.  Fringes due to Mechanism 1 turn out to be rare or nonexistent in high-mobility 2DEGs such as the one studied here due to the low density of hard scatterers \cite{Jura-NaturePhysics}.

Mechanism 2 for fringe formation is depicted in Fig. 1(c) and is relevant for our sample: electrons are emitted from the QPC, reflect off the tip, off the gates, off the tip again, and are finally retransmitted through the QPC (path depicted in red).  The path length difference between the red path and blue path is $2L$.  Therefore, for $L>L_T/2$, we expect the interference fringes due to mechanism 2 to fade.

In our sample $L_T/2 = 320 \ \mathrm{nm}$ at $T = 1.7 \ \mathrm{K}$ and $L_T/2 = 1.6 \ \mathrm{\mu m}$ at $T = 350 \ \mathrm{mK}$ (assuming a measured $\lambda/2 = 38 \ \mathrm{nm}$ as discussed below).  Figure 1(d) shows an image of electron flow taken at $1.7 \ \mathrm{K}$ in which fringes are not visible.  Mechanism 1 does not occur because there are no hard scatterers within the electron flow, and mechanism 2 is not seen because the imaging area is farther than $L_T/2$ from the QPC.  Figure 1(e) shows an image of electron flow taken over the same device but at $350 \ \mathrm{mK}$, and strong interference fringes are now visible in the bottom portion of the scan due to mechanism 2.  All images presented below were taken at $350 \ \mathrm{mK}$.  In other images we measure an average fringe spacing of $\lambda/2 = 38 \ \mathrm{nm}$ over many fringes far from the QPC.  Although the bulk electron density of the 2DEG suggests $\lambda_F/2 = 32 \ \mathrm{nm}$, we expect that the observed $\lambda/2$ will differ from the bulk value due to partial depletion of the 2DEG caused by the tip \cite{Topinka-Science,LeRoy-LocalDensity} and random local density variations \cite{LeRoy-LocalDensity}.  Near the QPC the local fringe spacing is further modified by partial depletion from the QPC gates and geometrical effects \cite{EPAPS}.  We assume an average $\lambda/2 = 38 \ \mathrm{nm}$ and note below when a local measurement of $\lambda/2$ is different from this average value.

\begin{figure}
\begin{center}
\includegraphics[width=3.0in]{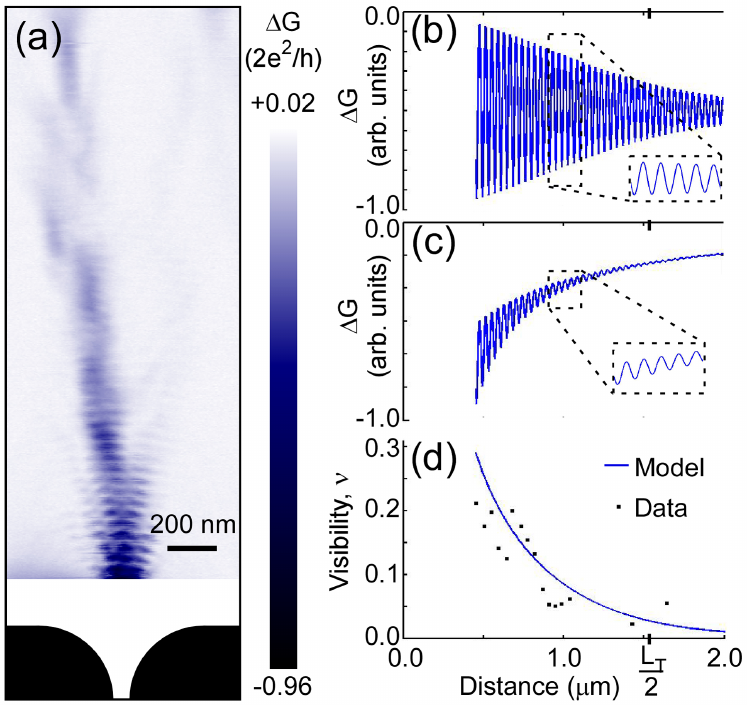}
\end{center}
\caption[Disappearance of interference fringes]{Disappearance of interference fringes.  (a) Image of electron flow showing strong interference fringes close to the QPC gates and a disappearance of fringes farther away.  (b) Idealized calculation of how interference visibility due to two paths of equal magnitude decreases farther from a coherent source due to thermal averaging.  (c) Calculation showing the disappearance of interference fringes due to two paths with fluxes decreasing as $1/L$ and $1/L^3$, respectively, as well as thermal averaging.  (d) Comparison of visibility $\nu$ for data in (a) and model in (c).}
\end{figure}

As expected, in Fig. 2(a) the interference fringe intensity fades with distance away from the QPC.  Figure 2(b) shows a calculation of how interference between two paths of equal magnitude fades over a characteristic distance $L_T/2 = 1.6 \ \mathrm{\mu m}$ due to thermal averaging.  This simple model underestimates the degree to which fringes fade as a function of distance from the QPC.  Electrons reflect off of the QPC and tip depletion regions with an angular spread; each reflection causes the current density to scale as $1/L$ at large distances $L$ \cite{Topinka-Thesis}.  Therefore, the fluxes of the blue and red paths in Fig. 1(c) should drop off roughly as $1/L$ and $1/L^3$ respectively.  Figure 2(c) shows a calculation of how interference fades taking into account thermal averaging and this rough distance dependence \cite{Normalize}.

Figure 2(d) shows how the visibility, $\nu =$(oscillation amplitude)/average, of interference fringes in our experimental data (black squares) compares to our calculation from Fig. 2(c) (blue line).  We extract $\nu$ from our data by fitting a sine curve to local cuts of interference fringes.  While the flux scattered from the tip varies non-trivially with position, the data follow the model well.  While a long $L_T$ (i.e. low temperature) is crucial to observing fringes caused by mechanism 2, geometrical attenuation of reflected waves ultimately limits the distance over which fringes are visible.

\begin{figure}
\begin{center}
\includegraphics[width=3.375in]{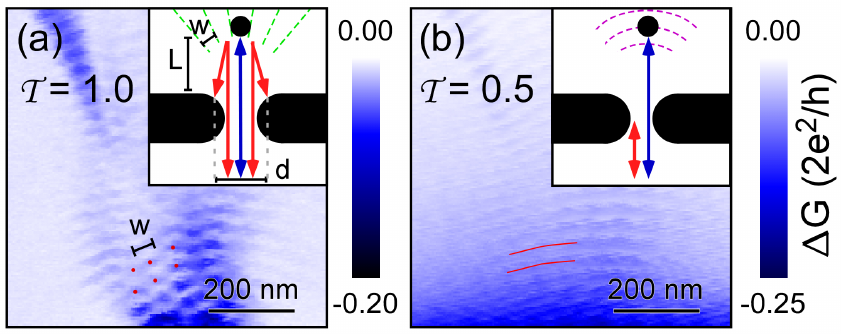}
\end{center}
\caption[Changing transmission of QPC]{Changing transmission of QPC.  (a) Image of electron flow at $\mathcal{T}=1.0$ showing a checkerboard interference pattern.  The centers of squares of the checkerboard are denoted with red circles.  Inset: schematic for mechanism causing checkerboard interference pattern where double-ended arrows indicate roundtrip paths.  The dashed green lines are lines along which the two red paths completely destructively interfere.  (b) Image of electron flow at $\mathcal{T}=0.5$ showing more of a ring pattern.  Red lines denote these rings.  Inset: schematic for mechanism causing ring interference pattern.  The dashed purple lines are those along which the blue path length is constant.}
\end{figure}

We next explore new interference patterns near the QPC and how the transmission coefficient $\mathcal{T}$ of the QPC affects the interferometer.  In Fig. 3(a), we image flow as before at $\mathcal{T}=1$ (i.e. $G=2e^2/h$), and in Fig. 3(b), we image flow at $\mathcal{T}=0.5$ (i.e. $G=e^2/h$).  At $\mathcal{T}=1$, we observe the interference forms a checkerboard pattern in certain areas (red circles in the center of ``squares'' are an aid to the eyes).  The interference pattern at $\mathcal{T}=0.5$ does not have as strong a checkerboard pattern and instead has a more ring like structure [denoted by red lines; see supplementary information \cite{EPAPS} for plots of the signal derivative and discussion of vibrations at the bottom of Fig. 3(b)].  

We understand the two different types of interference patterns observed in Fig. 3(a) and 3(b) as due to different sets of interfering paths.  In Fig. 3(a) at $\mathcal{T}=1$, interference happens as previously discussed, but we take into account reflections off both QPC gates because of the proximity to the QPC.  As depicted in Fig. 3(a) inset, the directly backscattered path (blue path) interferes with both paths reflecting off the QPC gates (red paths).  At certain distances from the QPC (green dashed lines), the two red paths destructively interfere.  As the tip moves along a green line, the blue path has no other path with which to interfere and we observe no fringes.  However, when the tip is between green lines, the summed red paths have a net amplitude, the magnitude of which changes sign on opposite sides of a green line.  The blue path interferes with the net red paths, and this creates a checkerboard pattern (see supplementary information \cite{EPAPS} for a predicted pattern).

The lateral spacing $w$ of squares in the checkerboard pattern allows us to estimate the distance $d$ between the reflection points of the red paths off the QPC gates as $w \approx \lambda \sqrt{L^2+(d/2)^2}/(2d)$ (both $w$ and $d$ denoted in Fig. 3(a) inset).  We measure a local fringe spacing of $\lambda/2 = 43 \ \mathrm{nm}$ here.  We find $w = 55 \ \mathrm{nm}$ and $L \approx 400 \pm 50 \ \mathrm{nm}$ (uncertainty is in the size of the tip-induced depletion disk and where electrons reflect off the QPC gates), giving $d \approx 340 \ \mathrm{nm}$.  This is in good agreement with the $d = 410 \ \mathrm{nm}$ predicted by a calculation of the gate potential by a three-dimensional Schr\"{o}dinger-Poisson solver (SETE code provided by M. Stopa \cite{Stopa-SETE}).

In Fig. 3(b) at $\mathcal{T}=0.5$, another interference mechanism shown in the inset becomes important.  Now, direct reflection from the QPC itself (red path) interferes with the path directly backscattered by the tip (blue path).  Rings of constant path length back to the QPC (dashed purple lines) are rings of constant phase for the blue path, and hence constant phase of interference in our measurement.  Indeed, we observe a stronger ring pattern in Fig. 3(b).

\begin{figure}
\begin{center}
\includegraphics[width=3.375in]{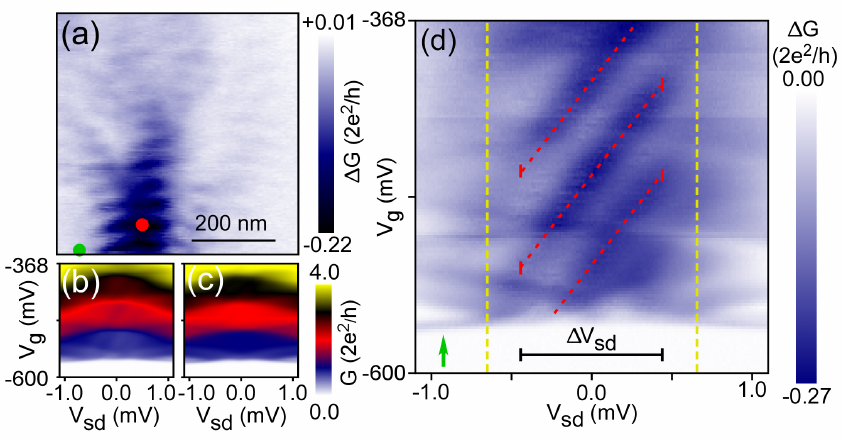}
\end{center}
\caption[Changing electron wavelength (energy) and cavity length]{Changing electron wavelength (energy) and cavity length.  (a) Image of electron flow.  Red circle denotes where the tip was placed when conductance was measured as a function of $V_g$ and $V_{sd}$ in (b), and green circle denotes the location of the tip for the measurements in (c).  The two locations are chosen to have similar capacitive coupling to the QPC.  (d) Subtraction of (c) from (b) shows features that are just due to the presence of the tip.  Red dashed lines denote the constant phase condition of the cavity when changing the electron wavelength ($V_{sd}$) and cavity length ($V_g$).  The dashed yellow lines show the voltage at which we calculate electron-electron scattering to cause a reduction in interference visibility by half.  The green arrow points to diamondlike features which are an artifact of the subtraction; (b) and (c) have been shifted slightly in $V_g$ due to different capacitive coupling from the tip.}
\end{figure}

We next study changes in cavity length and injected electron wavelength (i.e. energy) and how finite injection energy leads to dephasing.  With the tip positioned in the electron flow in Fig. 4(a) (at the red circle) $700 \ \mathrm{nm}$ from the QPC, we vary the gate voltage on the QPC, $V_g$, and source-drain bias across the QPC, $V_{sd}$.  Setting $V_g$ more negative widens the depletion region underneath the gate, effectively decreasing the distance $L$.  A negative $V_{sd}$ injects electrons at higher energies and shorter wavelengths.  Figure 4(b) shows the conductance, $G$, with the tip in the electron flow.  The prominent diamondlike features stem from quantized conductance of subbands in the QPC \cite{Kouwenhoven-QPCDiamond}.  To determine the effects of the tip, in Fig. 4(c) we take the same conductance measurements with the tip out of the electron flow [at the green circle in Fig. 4(a)].  We subtract the data in Fig. 4(c) from those in Fig. 4(b) to produce Fig. 4(d), a plot of $\Delta G$, which shows just the effect of the tip.

In Fig. 4(d), strong diagonal features (marked with red dashed lines) indicate the constant phase condition for the interferometer $2kL = 2 \pi n$, where $k$ is the wave number and $n$ is an integer.  The spacing between red dashed lines along the $V_g$ axis corresponds to changing the cavity length $L$ by $\lambda/2$.  The interval between red dashed lines along $V_g$ is $65 \ \mathrm{mV}$, implying that increasing the gate voltage extends the depletion region laterally at a rate of $0.6 \ \mathrm{nm}/\mathrm{mV}$, which agrees with predictions from our SETE calculations.  Additionally, the spacing between red dashed lines along the $V_{sd}$ axis should correspond to changing the electron wavenumber $k$ by $\pi/L$ inside the cavity of fixed length $L$.  We measure that the spacing between red dashed lines along $V_{sd}$ is $420 \ \mathrm{\mu V}$ and predict, based on our estimate of $L$ from SETE calculations, that the spacing should also be $420 \ \mathrm{\mu V}$.

The dependence of interference fringes on $V_{sd}$ provides a useful technique for spatially studying dephasing.  In Fig. 4(d), we see that at high $|V_{sd}|$ the diagonal features (marked with red dashed lines) disappear.  This can be explained by inelastic scattering that causes electrons to lose phase information and no longer interfere.  To quantify the energy dependence, for each $V_{sd}$ we fit a sine curve to $\Delta G$ oscillations as a function of $V_g$ and use the amplitude as a measure of interference strength.  Although the QPC adds non-trivial subband features to $\Delta G$, we estimate the 2 values of $V_{sd}$ at which the interference strength is reduced to $50 \%$ of its maximum [denoted by the end of the red dashed lines with short vertical lines in Fig. 4(d)] and find their difference $\Delta V_{sd} = 920 \ \mathrm{\mu V}$.

Electron-electron scattering increases at high-bias and is generally the dominant source of inelastic scattering in this regime \cite{Yacoby-DoubleSlit}.  We calculate the expected $\Delta V_{sd}$ from the electron-electron scattering length $L_{e-e}$ \cite{Giuliani} [the average path length of electrons $3L = \ln(2) L_{e-e}(V_{sd})$].  Using the density measured from the average fringe spacing, the expected $\Delta V_{sd}$ is $1250 \ \mathrm{\mu V}$ [yellow dashed lines in Fig. 4(d)], close to our data.  In the future a more detailed spatial dependence study may give insight into dephasing mechanisms.

We have demonstrated a thorough understanding of an electronic interferometer using an SGM tip and QPC.  We show new ways of measuring the spatial dependence of dephasing, which can be used to probe electron-electron scattering in a 2DEG.  Additionally, this interferometer may prove useful for studying electron interactions and complex flow patterns in the ``0.7 structure'' regime \cite{Thomas-0.7} of a QPC in which transport measurements indicate correlated electron structure.  Thus we have investigated changes in transport through the interferometer as a function of QPC transmission.\newline

We thank A. Sciambi for help characterizing the 2DEG sample.  We are grateful to M. Stopa and NNIN/C for making SETE available.  This work was supported by the Stanford-IBM Center for Probing the Nanoscale, an NSF NSEC (Grant No. PHY-0425897).  Work was carried out in part at the Stanford Nanofabrication Facility of NNIN supported by NSF (Grant No. ECS-9731293).  M.P.J. acknowledges support from the NDSEG program.  D.G.-G. recognizes support from the Packard Foundation.\small\newline

\noindent$^*$ Present address: Hitachi GST, San Jose, California 95135

\noindent$^{\dag}$ Corresponding author; goldhaber-gordon@stanford.edu \normalsize

\newcommand{\noopsort}[1]{} \newcommand{\printfirst}[2]{#1}
  \newcommand{\singleletter}[1]{#1} \newcommand{\switchargs}[2]{#2#1}

\newpage

\setlength{\topmargin}{-0.5in}
\setlength{\oddsidemargin}{-1in}
\setlength{\evensidemargin}{-1in}

\begin{figure}[t]
\begin{center}
\includegraphics{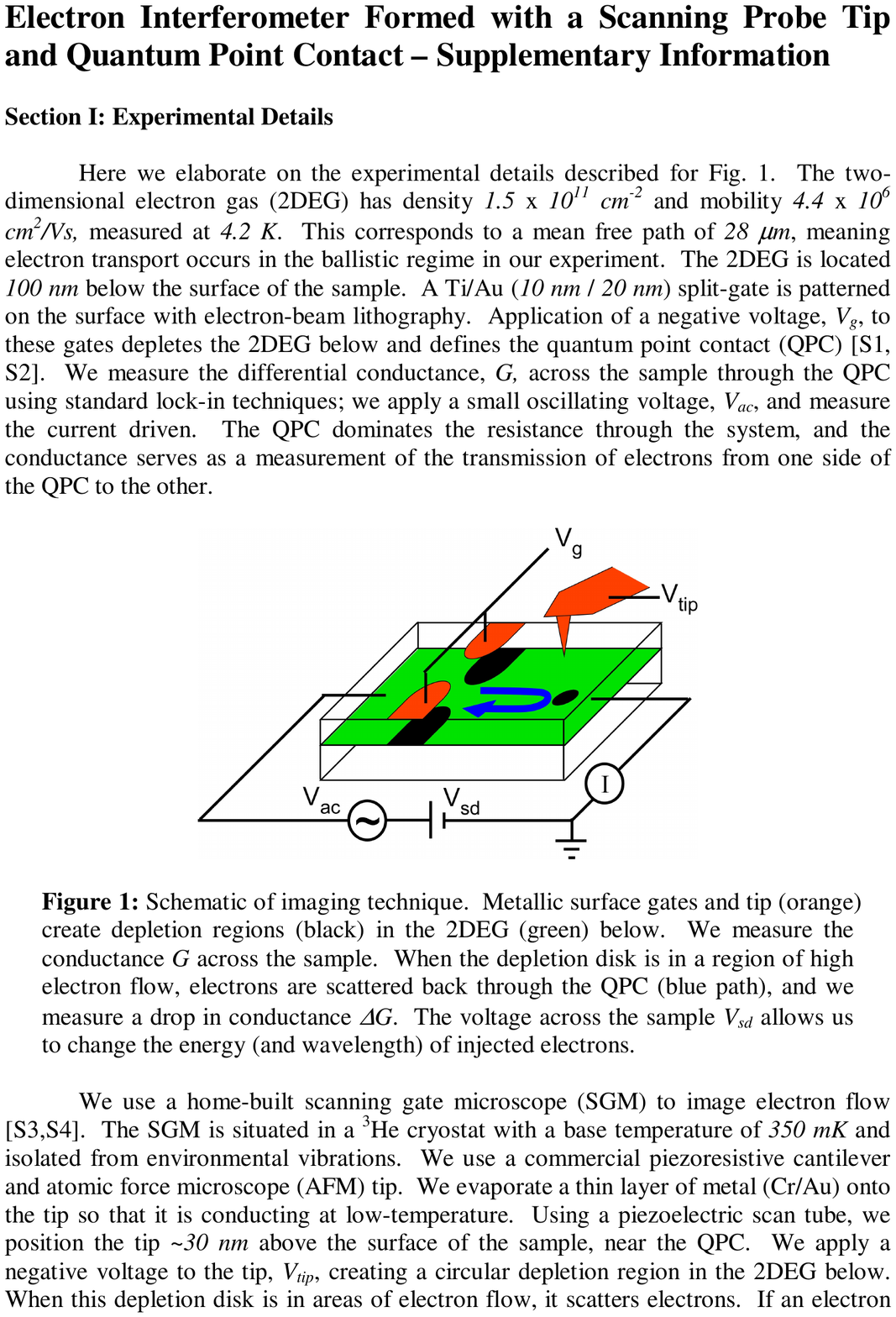}
\end{center}
\end{figure}

\begin{figure}[t]
\begin{center}
\includegraphics{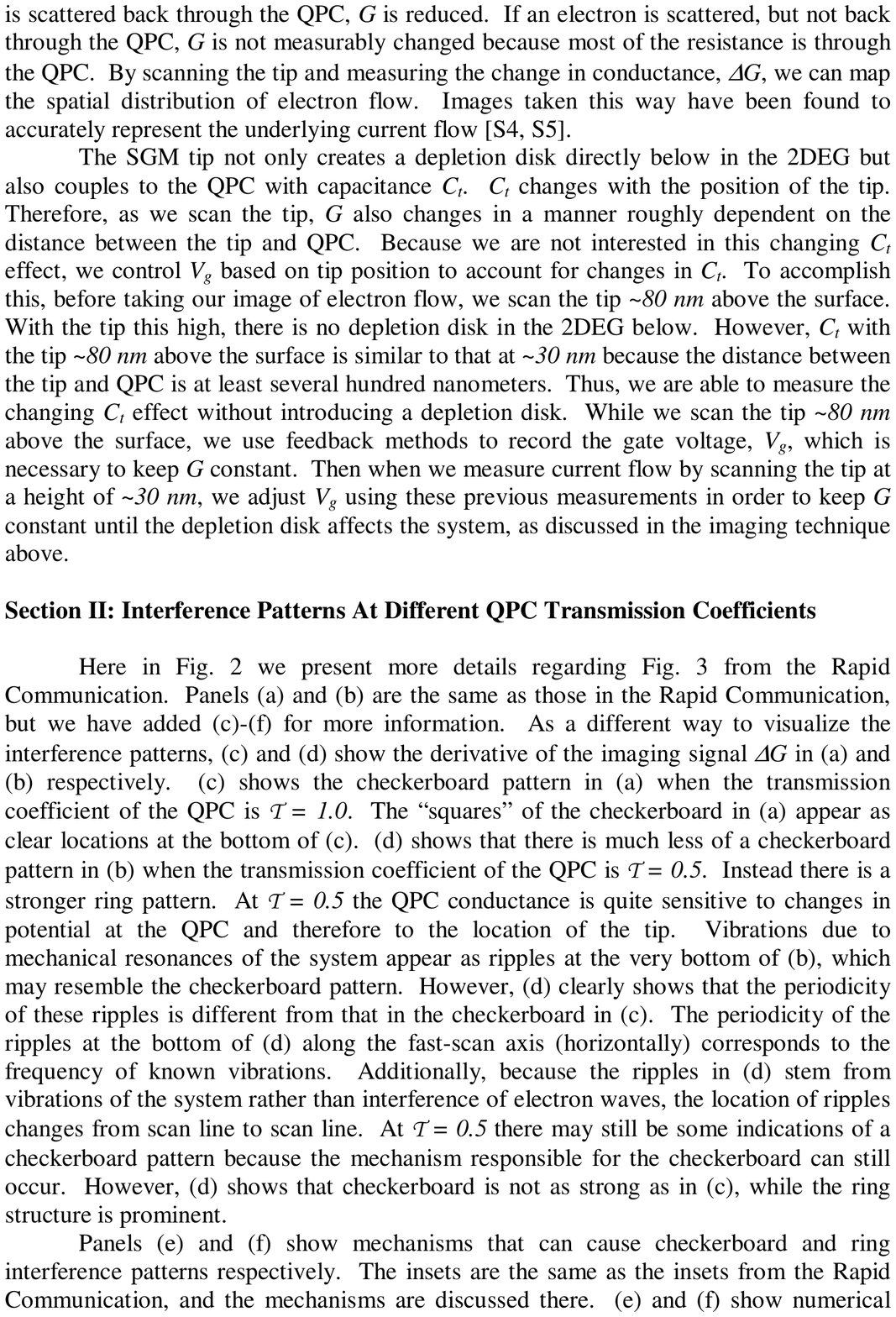}
\end{center}
\end{figure}

\begin{figure}[t]
\begin{center}
\includegraphics{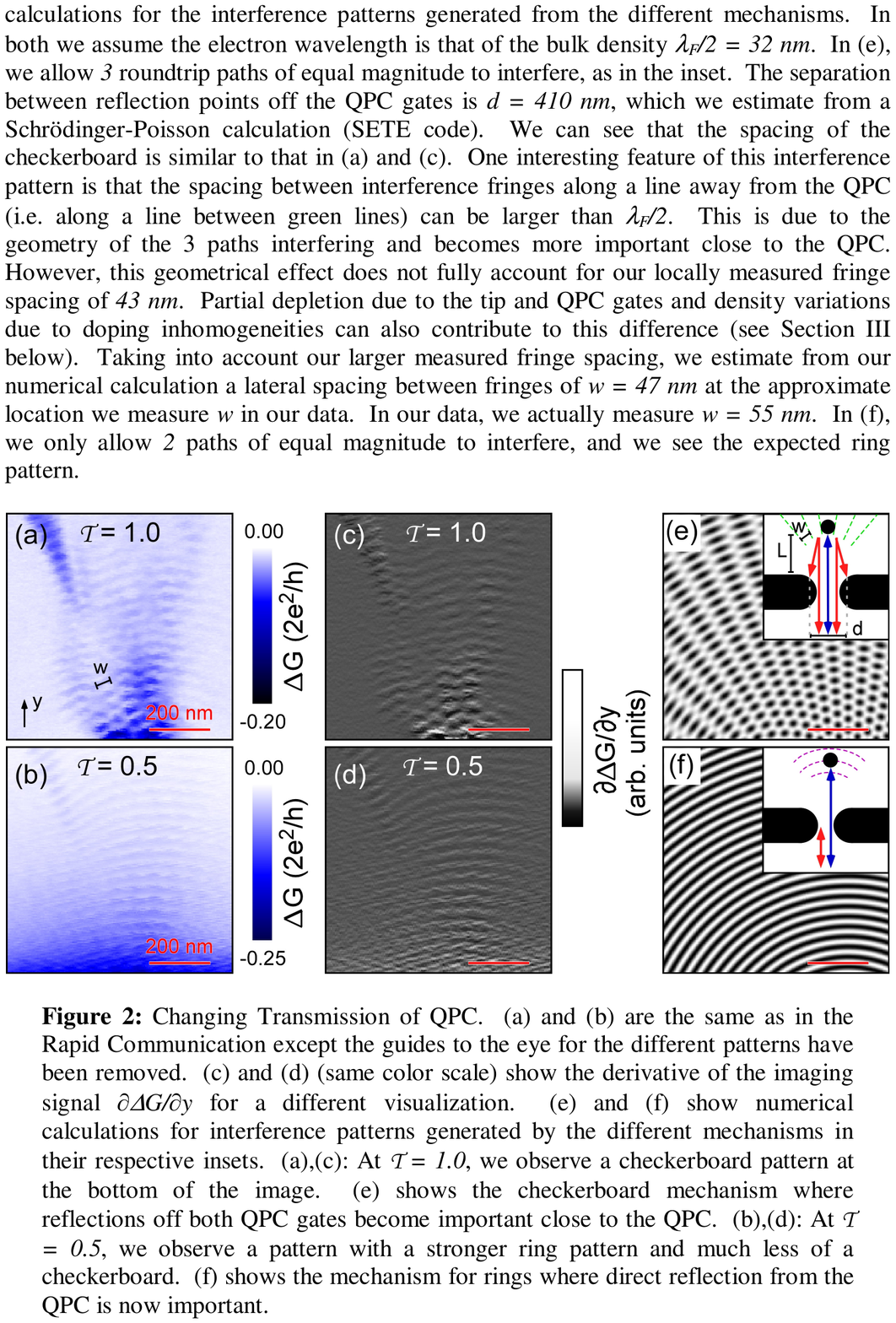}
\end{center}
\end{figure}

\begin{figure}[t]
\begin{center}
\includegraphics{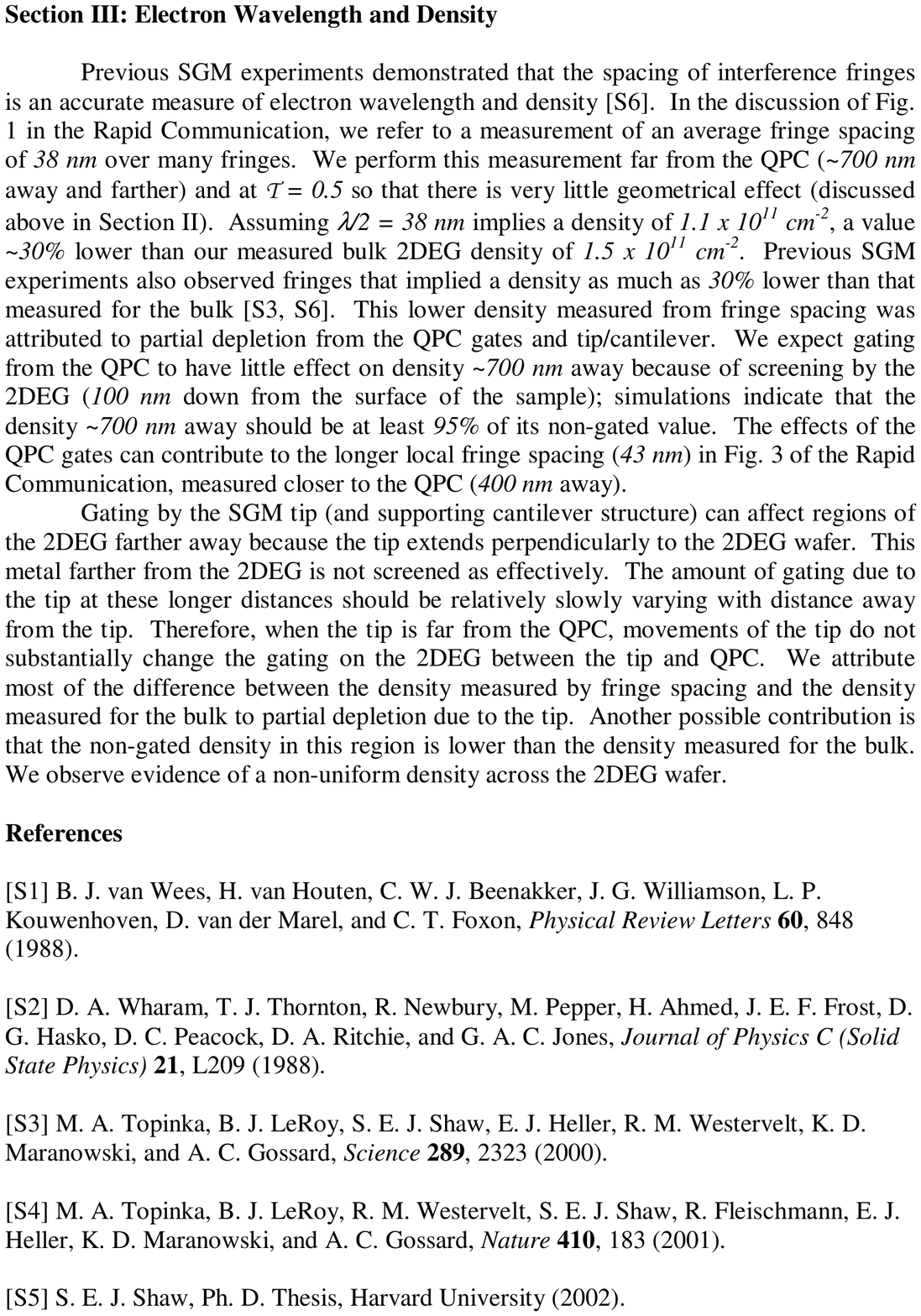}
\end{center}
\end{figure}

\begin{figure}[t]
\begin{center}
\includegraphics{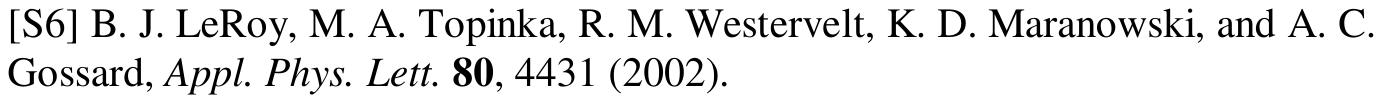}
\end{center}
\end{figure}


\begin{thebibliography}{10}
\bibitem{Normalize}
The fluxes of the 2 paths are normalized by $\lambda_F$ and $\lambda_F^3$ to make them dimensionless.  No parameters are fit.
\end{thebibliography}
\end{document}